\documentstyle[epsfig,12pt]{article}
\title{
\begin{flushright}
{\normalsize NUC-MINN-00/19-T \\}
\end{flushright}
\vspace*{0.3in}
{\bf Will Strangeness Win the Prize?}}
\author{{\bf Joseph I. Kapusta}\vspace*{0.1in}\\
 {\it School of Physics and Astronomy, University of Minnesota}\\
 {\it Minneapolis, MN 55455}}

\date{}

\begin{document}

\maketitle

\begin{center}
Abstract
\end{center}

Five groups have made predictions involving the production of strange hadrons 
and entered them in a competition set up by Barbara Jacak, Xin-Nian Wang and 
myself in the spring of 1998 for the purpose of comparing to first year
physics results from RHIC.  These predictions are summarized and evaluated.

\section{Introduction}

In the spring of 1998 Barbara Jacak, Xin-Nian Wang and I organized a workshop 
named {\it Probes of Dense Matter in Ultrarelativistic Heavy Ion Collisions} 
at the Institute for Nuclear Theory (INT), University of Washington, 
Seattle.  During this workshop we discussed the desirability of making 
predictions for the first data to come from RHIC rather than fitting model 
parameters after the data was taken.  To encourage these predictions we set up
a competition.  The winner would be the one that made the best prediction for
any data taken during the first year of running.  We three would be the 
judges.  The winner would be awarded a case of wine that now resides in a
cellar in Washington.  This competition was widely advertised by the 
organizers, including announcements made at the last Quark Matter 
meeting and in a Division 
of Nuclear Physics, American Physical Society newsletter.  The deadline for 
submissions is now past since the first data at RHIC has been taken.

Eleven submissions were made to the competition.  These may be viewed at 
the web site {\em ftp://www-hpc.lbl.gov/} by clicking on {\em int-prediction}.
Of the eleven, five involve predictions for strangeness production of one sort
or another.  These may be categorized as phase space models, hydrodynamic 
models, and cascade or transport models.  I will briefly review the 
predictions here.  For more details the interested reader is directed to the
aforementioned web site.  All predictions are for Au+Au collisions at
the full RHIC beam energy of 100 GeV.  Unless otherwise stated the 
predictions are for central collisions.

\section{Phase Space Model}

Given the number of people who have used the thermal/statistical/phase space 
model over the years it is somewhat surprising that there is only one entry in 
this category.  This entry is from J. Rafelski and J. Letessier.  In its pure 
form, with kinetic and chemical equilibrium, one calculates the abundances of 
hadrons of any type using a Fermi-Dirac or Bose-Einstein phase space 
distribution.  One may allow for a system in kinetic, but not chemical, 
equilibrium by introducing a factor $\gamma$ which multiplies the phase 
space density.  Then the parameters consist of temperature $T$; baryon,
electric charge, and strangeness chemical potentials $\mu_i$ 
(or fugacity $\lambda_i = \exp(\mu_i/T)$); and a factor of $\gamma_q$ for 
$u$ and $d$ and $\gamma_s$ for 
$s$ quark content, respectively.  Since this approach has no nucleus-nucleus 
dynamics, one must estimate these parameters at the moment the matter ceases 
interaction, or freezes out.  Of course, they can be fit to the data once it 
becomes available.  Based on existing data from the SPS and elsewhere, and 
making reasonable and judicious assumptions about what happens at higher 
energies, the predictions are shown in Tables I and II.
Various possibilities for fugacities and 
enhancement factors are given, but in all cases the temperature is taken to be 
150 MeV.  It is the numerical value of $T$ and the fact that the $\gamma$ are 
not equal to 1 that primarily differentiate this approach from other 
thermal model approaches.  Actually there is not much room to maneuver 
if one takes the fitted values at the SPS 
($T = 160 \pm 10$ and $\mu_B = 200 \pm 70$ MeV) and 
extrapolates to RHIC.  (See the presentation by J. Cleymans in this 
proceedings.)  If the RHIC data show anything close to the predictions of the 
phase space model as they do at the SPS, particularly for strange hadrons, one 
is still confronted by the details of the hadronization dynamics that somehow 
populate phase space.
 
\section{Hydrodynamic Models}

Hydrodynamic models of high energy collisions go back to Landau in the 1950's. 
The advantage of this approach is that it readily incorporates the equation of 
state and a possible phase transition which is, after all, what we are 
primarily interested in.  The main issues: What are the initial conditions?
How accurate is the assumption of local thermal and chemical equilibrium?  
How is the final stage from hydrodynamic flow to free streaming 
particles handled?  Two groups have entered the competition based on the 
hydrodynamic approach.

A. Dumitru and D. Rischke solved the relativistic perfect fluid equations 
assuming longitudinal scaling flow.  They used various microscopic models, 
namely PCM, RQMD, FRITIOF, and HIJING, to 
assist in estimating the initial conditions.  The baryon rapidity density 
$dN_B/dy$ was taken to be 25 and the entropy per baryon was taken to be 200.  
The formation time of the system was taken as 0.6 fm/c, which is about
1/10 of the radius of a gold nucleus.  This resulted in an 
initial energy density of 
17 GeV/fm$^3$ and baryon density of 2.3 times normal nuclear matter density.
For a near ideal quark-gluon plasma this yields an initial temperature of 
300 MeV and a light quark chemical potential of 47 MeV.  
The phase transition to hadrons is first order with a critical temperature of
160 MeV (when the baryon chemical potential is zero).  
The sensitivity of the observed spectra to several different initial 
density profiles was investigated.  They found that the average transverse 
flow velocity is rather similar to that in central Pb+Pb collisions at 
the SPS due to the stall of the flow within the mixed phase.  
Figure 1 shows the calculated transverse momentum spectra for 11 
different species of hadrons.  The left panel is the distribution immediately 
after the phase transition is completed, the right panel shows the 
distribution calculated along the isotherm of $T = 130$ MeV.  One sees 
clearly that the flow continues to evolve after the phase transition.  
It is especially apparent for heavier hadrons.  Figure 2 
shows average transverse velocity (left panel) and transverse momentum (right 
panel) as a function of the mass of the hadron.  Hydrodynamics predicts a 
linear increase of the average transverse momentum with mass, whereas 
FRITIOF alone shows a more complicated dependence on the flavor 
composition of the hadrons.  As can be seen, the results do depend on the 
hypersurface on which the hadron spectra are computed.

S. Bass and A. Dumitru also solved boost invariant hydrodynamic equations.
The initial conditions were essentially identical to those of 
Dumitru and Rischke.
The initial transverse energy and baryon density profile was taken to be 
proportional to $\sqrt{1-(r_T/R_T)^2}$ with the transverse radius $R_T = 6$ fm 
and the formation time 1/10 of that.  After the phase transition is complete 
and the matter has been evolved to a temperature of 130 MeV the hadron 
spectra are computed using the Cooper-Frye formula, as in Dumitru and Rischke,
but the resulting spectra is then used as an initial condition for the 
hadronic cascade UrQMD.  In this way the transition from perfect local 
thermal equilibrium to free streaming 
hadrons is done in a more sophisticated manner than making a sudden, abrupt 
transition from one to the other.  The resulting transverse mass distributions 
for baryons with different strangeness content and for pions and kaons are 
shown in Figure 3 for SPS, RHIC, and LHC energies.  At least for the SPS 
and RHIC there is not much effect of using UrQMD as the intermediary 
between hydrodynamic flow and free streaming.  Figure 4 shows the 
average transverse momentum for 
hadrons as a function of their mass for the three energies.  Once again the 
effect of UrQMD is small but measurable.

These entries nicely illustrate the beautiful evolution of a quark-gluon 
plasma, through a phase transition, into a hadronic gas and the final 
nonequilibrium freezeout into free streaming particles.  They could be 
elaborated even more with the use of imperfect fluid dynamics 
(viscosity and heat conduction) and finite nucleation rates for the 
phase transition.  The main uncertainty is still the choice of initial 
conditions.  The most reasonable choices are to take them 
from a microscopic model of the early stages of a heavy ion collision, such as 
these two entries do, or to adjust them so that the computed final state 
spectra agree with data.  Although the second choice is certainly valid and 
worthwhile it is difficult to refer to it as a prediction.  

\section{Microscopic Transport Models}

The most ambitious entries compute the whole nucleus-nucleus collision from 
beginning to end on the basis of microscopic dynamics with no assumptions 
about local equilibrium.  By necessity this is done with Monte Carlo 
techniques.  There are two entries in this category which are relevant for 
strange particle production.

B. Zhang, C.M. Ko, B.-A. Li and Z. Lin take the initial 
parton distributions from HIJING and 
evolve them according to their parton cascade model ZPC.  They allow the 
partons to hadronize, after which hadrons interact according to the cascade 
model ART until all particles free stream to infinity.  As with any parton 
cascade there are uncertainties arising from the nature of infrared 
singularities in QCD.  The predictions for $K^+$ and for all mesons are 
shown in Figure 5.  The main point 
is that multiple collisions in the partonic and hadronic phases produce more 
particles than that input from HIJING.  The extra particle production 
occurs in the central rapidity range where the initial particle 
density is the highest, of 
course.  The kaon abundance is especially enhanced, by about 50\%.

The UrQMD collaboration entry was submitted by M. Bleicher.  The UrQMD creates 
particles at high energies from strings, and at low energies exclusive 
reactions are parametrized, usually in terms of resonances.  
It is a very ambitious model which can predict almost anything desired 
in a heavy ion collision.  This also means that it is easier to find 
fault with it, either in terms of the details of its 
dynamics, or in terms of predictions not consistent with experimental data.  I 
have selected a subset of all the submitted predictions involving strange 
hadrons.  In Figure 6 are shown the predictions for the rapidity distributions 
of protons, antiprotons, negative pions, and kaons.  In each panel there 
are two sets of numbers: the open symbols have meson-meson and meson-baryon 
and certain quark interactions turned off, the full symbols have them all 
turned on.  Just like the previous cascade approach, multiple scattering 
among the produced 
particles creates more particles.  It also shows that baryons are moved in 
toward central rapidity from the fragmentation regions.  Figure 7 shows the 
average transverse momentum for pions, kaons, and protons as functions of 
rapidity.  Multiple scatterings increase the mean transverse momentum for 
protons and kaons, but slightly reduce it for pions.  
This may be a consequence 
of the ability of higher energy pions to produce kaons whereas the 
lower energy pions remain.  UrQMD is even so bold as to predict the 
directed and elliptic flow parameters, $v_1$ and $v_2$, for every species 
of hadron.  These are shown for pions and kaons in Figures 8 and 9.  
Pions and kaons flow in about the same way, but the magnitude of the 
flow coefficients is greater for pions.  Similar figures comparing protons, 
antiprotons, and lambdas may be found at the web site.  These observables 
give information on the degree of equilibration and on the amount of 
energy loss of the hadrons as they traverse hadronic matter.

Any cascade-like dynamical model requires detailed assumptions about 
production cross sections, off-shell behavior of propagating particles, 
formation times, etc.  It is very good to see these firm predictions for 
they can be tested in the very near future!       

\section{Summary}

There is a proverb which says ``May you live in interesting times."  
The three of us eagerly look forward to comparing all the predictions 
with RHIC data as they emerge.  We should be ready to announce the 
winner in the summer of 2001.  The 
most exciting outcome would be something that was not predicted!  

\section*{Acknowledgements}  

This work was supported by the US Department of Energy under grant
DE-FG02-87ER40328.

\newpage

\begin{table}[t]
\caption{\label{table1}
For $\gamma_{s}=1.25,\,\lambda_{s}=1$ and $\gamma_{q}$, $\lambda_{q}$ as shown.
Top portion: strangeness per baryon $s/B$, 
energy per baryon $E/B$[GeV]  and  entropy per baryon $S/B$. Bottom portion:
sample of hadron ratios expected at RHIC.  From Rafelski and Letessier.}
\small
\vspace*{-0.2cm}
\begin{center}
\begin{tabular}{l|lllll}
$\gamma_{q}$                           & 1.25 & 1.5  &  1.5  &  1.5  & 1.60 \\
$\lambda_{q}$                          & 1.03 & 1.025&  1.03 & 1.035 & 1.03 \\
\hline
$E/B$[{\small GeV}]                    & 117  & 133  &  111  &  95   & 110 \\
$s/B$                                  & 18   & 16   &  13   &  12   & 12 \\
$S/B$                                  & 630  & 698  &  583  & 501   & 571 \\
\hline\hline
$p/{\bar p}$                           & 1.19 & 1.15 & 1.19  &  1.22 & 1.19 \\
$\Lambda/p$                            & 1.74 & 1.47 & 1.47  &  1.45 & 1.35 \\
${\bar\Lambda}/{\bar p}$               & 1.85 & 1.54 & 1.55  &  1.55 &1.44 \\
${\bar\Lambda}/{\Lambda}$              & 0.89 & 0.91 &  0.89 &  0.87 & 0.89 \\
${\Xi^-}/{\Lambda}$                    & 0.19 & 0.16&  0.16  &  0.16 & 0.15 \\
${\overline{\Xi^-}}/{\bar\Lambda}$     & 0.20 & 0.17 &  0.17 &  0.17 & 0.16 \\
${\overline{\Xi}}/{\Xi}$               & 0.94 & 0.95 &  0.94 &  0.93 & 0.94 \\
${\Omega}/{\Xi^-}$                     & 0.147&0.123 &  0.122&  0.122& 0.115 \\
${\overline{\Omega}}/{\overline{\Xi^-}}$
                                       & 0.156& 0.130&  0.130&  0.131& 0.122 \\
${\overline{\Omega}}/{\Omega}$         &  1   & 1   &  1   &  1   & 1   \\
$(\Omega+\overline{\Omega})/(\Xi^-+\overline{\Xi^-})$
                                       & 0.15 & 0.13 &  0.13 &  0.13 & 0.12 \\
$(\Xi^-+\overline{\Xi^-})/(\Lambda+\bar{\Lambda})$
                                       & 0.19 & 0.16 &  0.16 &  0.16 & 0.15 \\
${K^+}/{K^-}$                          & 1.05 & 1.04 &  1.05 &  1.06 & 1.05 \\
\end{tabular}
\end{center}
\vspace*{-.6cm} 
\end{table}

\begin{table}[htb]
\caption{\label{table2} $dN/dy|_{\mbox{\scriptsize central}}$ 
assuming $dp/dy|_{\mbox{\scriptsize central}}=25$.
From Rafelski and Letessier.}
\vspace*{-.2cm} 
\begin{center}
\begin{tabular}{ll|cccccccccc}
$\gamma_{q}$& $\lambda_{q}$  
& $b$ &  $p$ & $\bar p$ & $\!\!\Lambda\!\!+\!\!\Sigma^0\!\!$ &
$\!\!\overline{\Lambda}\!\!+\!\!\overline{\Sigma}^0\!\!$
&$\Sigma^{\pm}$&$\overline{\Sigma}^{\mp}$ & 
$\Xi^{^{\underline{0}}}$ &$\overline{\Xi}^{^{\underline{0}}}$&
 $\Omega\!=\!\overline\Omega$  \\
\hline
1.25& 1.03 & 17 & 25$^*$& 21 & 44 & 39 & 31 & 27 & 17 & 16 & 1.2  \\ 
1.5 & 1.025 & 13 & 25$^*$& 22 & 36 & 33 & 26 & 23 & 13 & 11 & 0.7 \\ 
1.5& 1.03 & 16 & 25$^*$& 21 & 37 & 33 & 26 & 23 & 12 & 11 & 0.7 \\ 
1.5 & 1.035 & 18 & 25$^*$& 21 & 36 & 32 & 26 & 22 & 11 & 10 & 0.7 \\ 
1.60& 1.03  & 15 & 25$^*$& 21 & 34 & 30 &24 &21 & 10 & 9.6 & 0.6 \\ 
\end{tabular}
\end{center}
\vspace*{-0.6cm} 
\end{table}

\begin{figure}
\centerline{\epsfig{figure=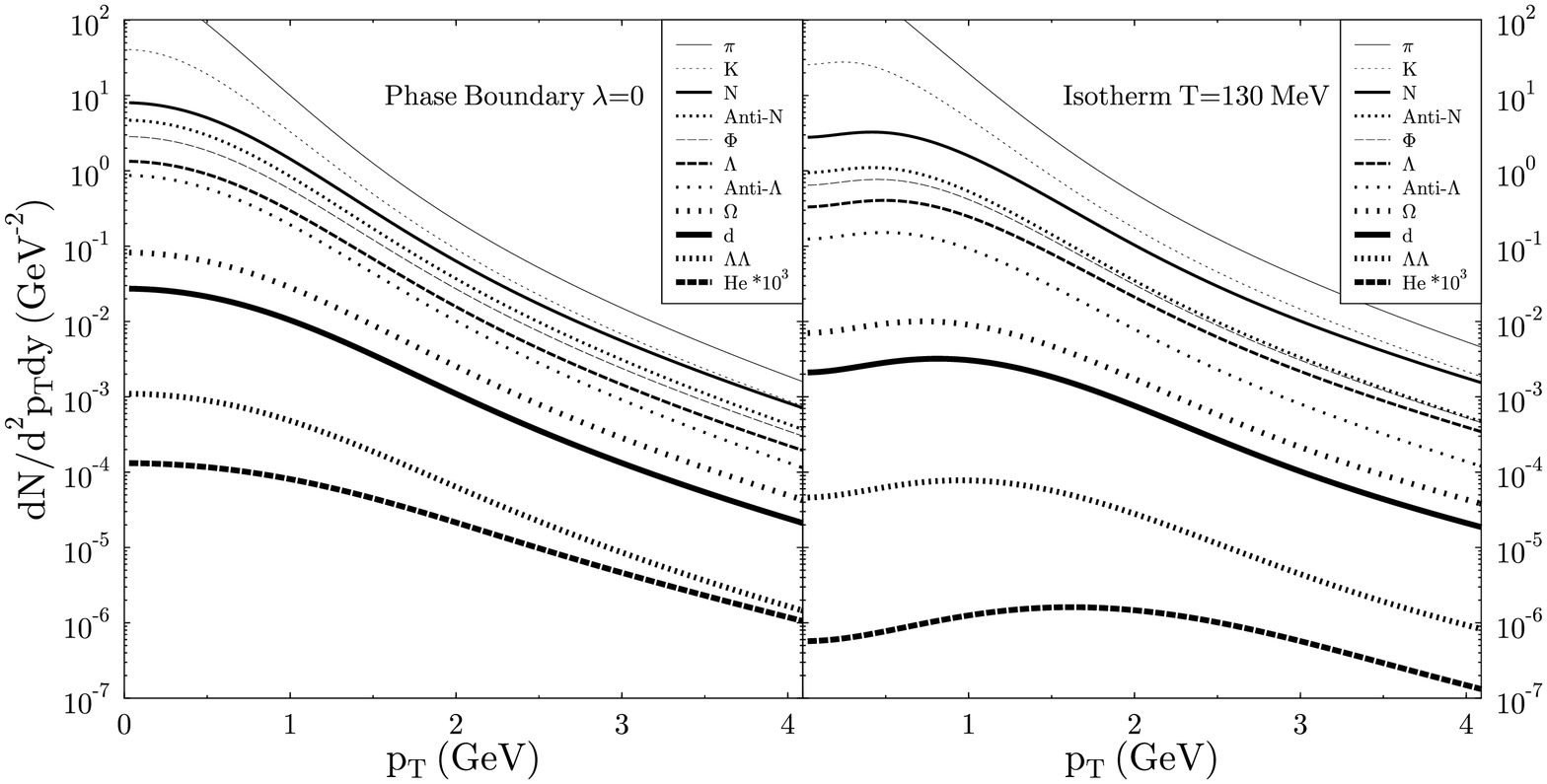,width=6.0in}}
\caption{Transverse momentum distribution at midrapidity.  Left panel is
immediately after completion of phase transition, right panel
is at the isotherm of 130 MeV.  From Dumitru and Rischke.}
\end{figure}

\begin{figure}
\centerline{\epsfig{figure=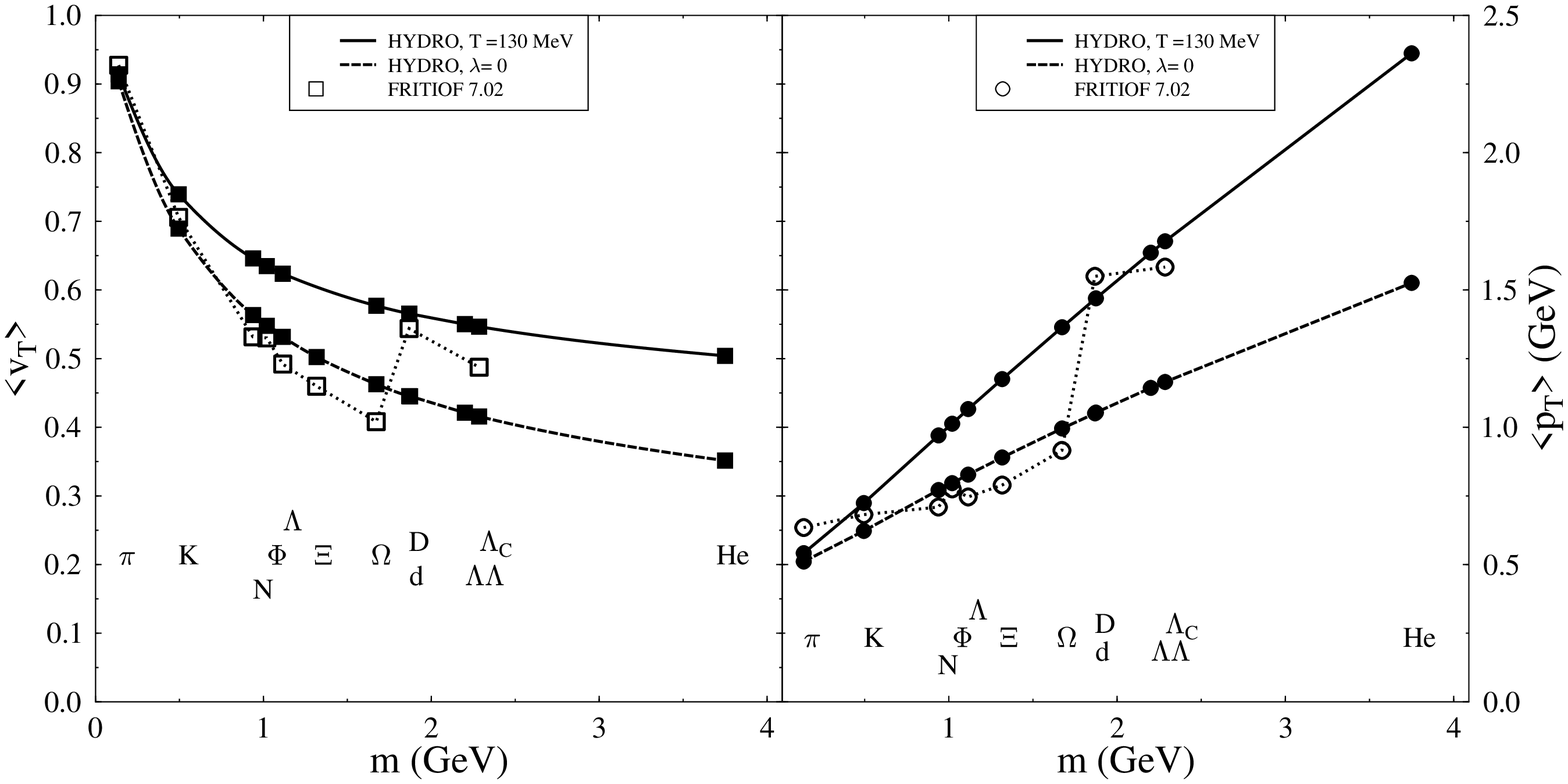,width=6.0in}}
\caption{Average transverse velocity and momentum of various hadrons
computed immediately after the phase transition (labeled $\lambda = 0$)
and at the isotherm of 130 MeV.  From Dumitru and Rischke.} 
\end{figure}

\begin{figure}
\centerline{\epsfig{figure=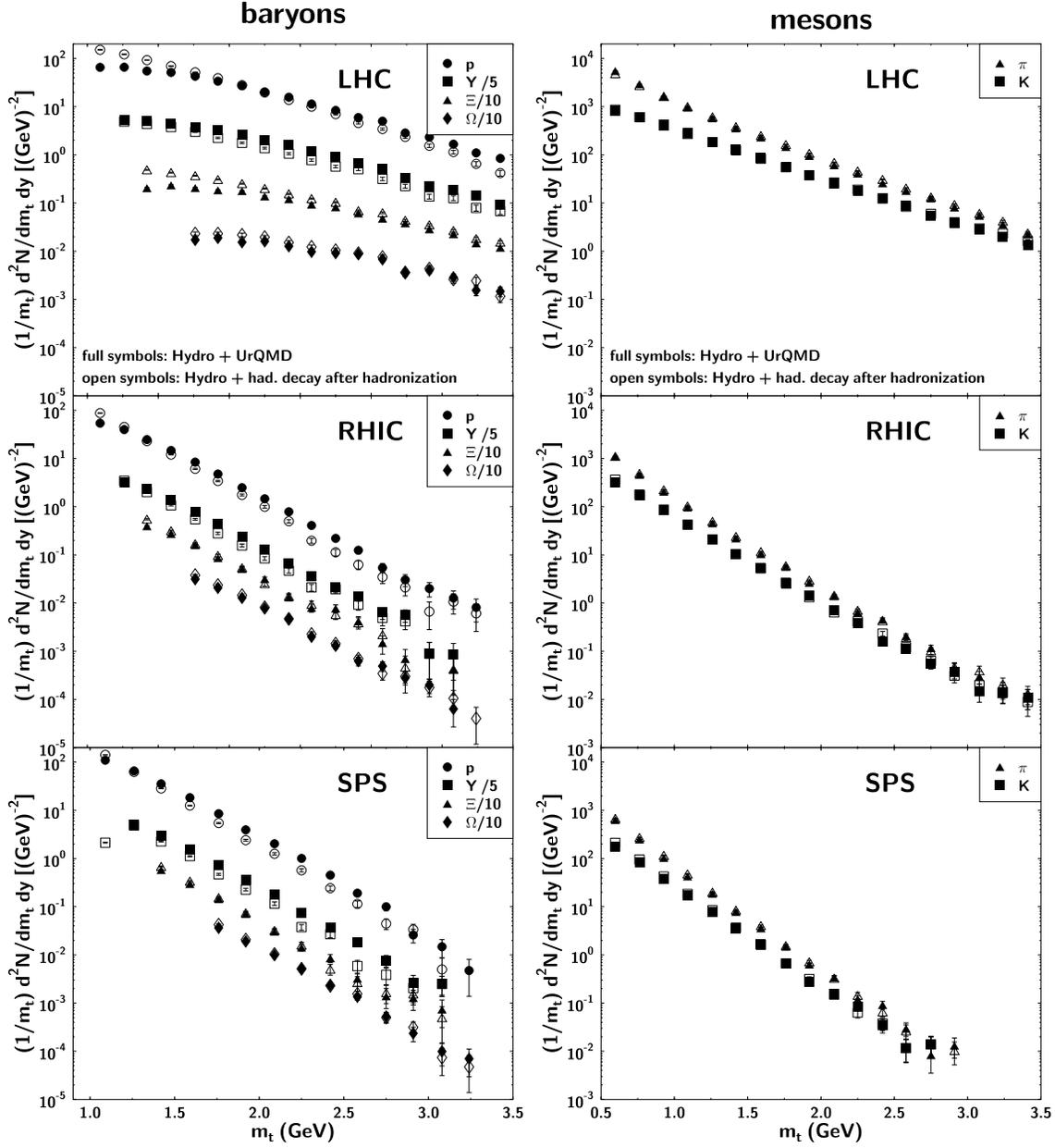,width=6.0in}}
\caption{Transverse mass distributions for central collisions. 
From Bass and Dumitru.}
\end{figure}
\begin{figure}

\centerline{\epsfig{figure=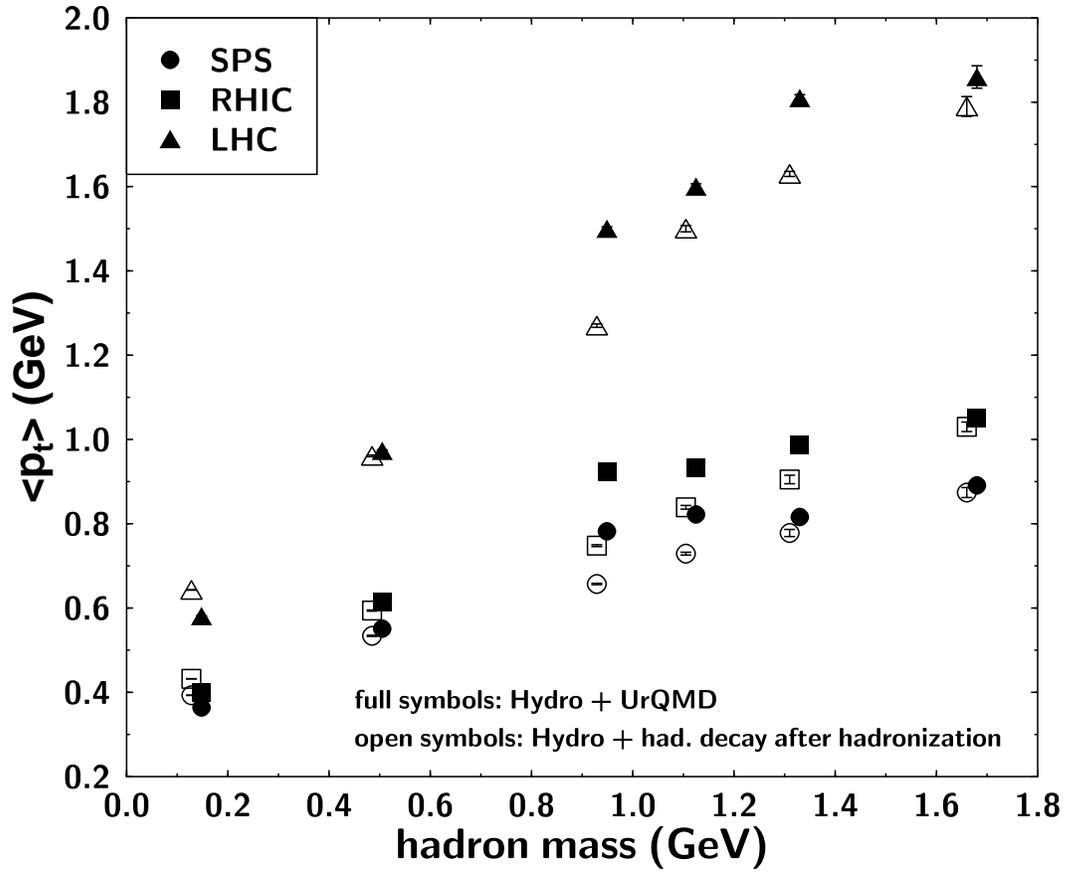,width=6.0in}}
\caption{Average transverse momentum for central collisions.  
From Bass and Dumitru.} 
\end{figure}

\begin{figure}
\centerline{\epsfig{figure=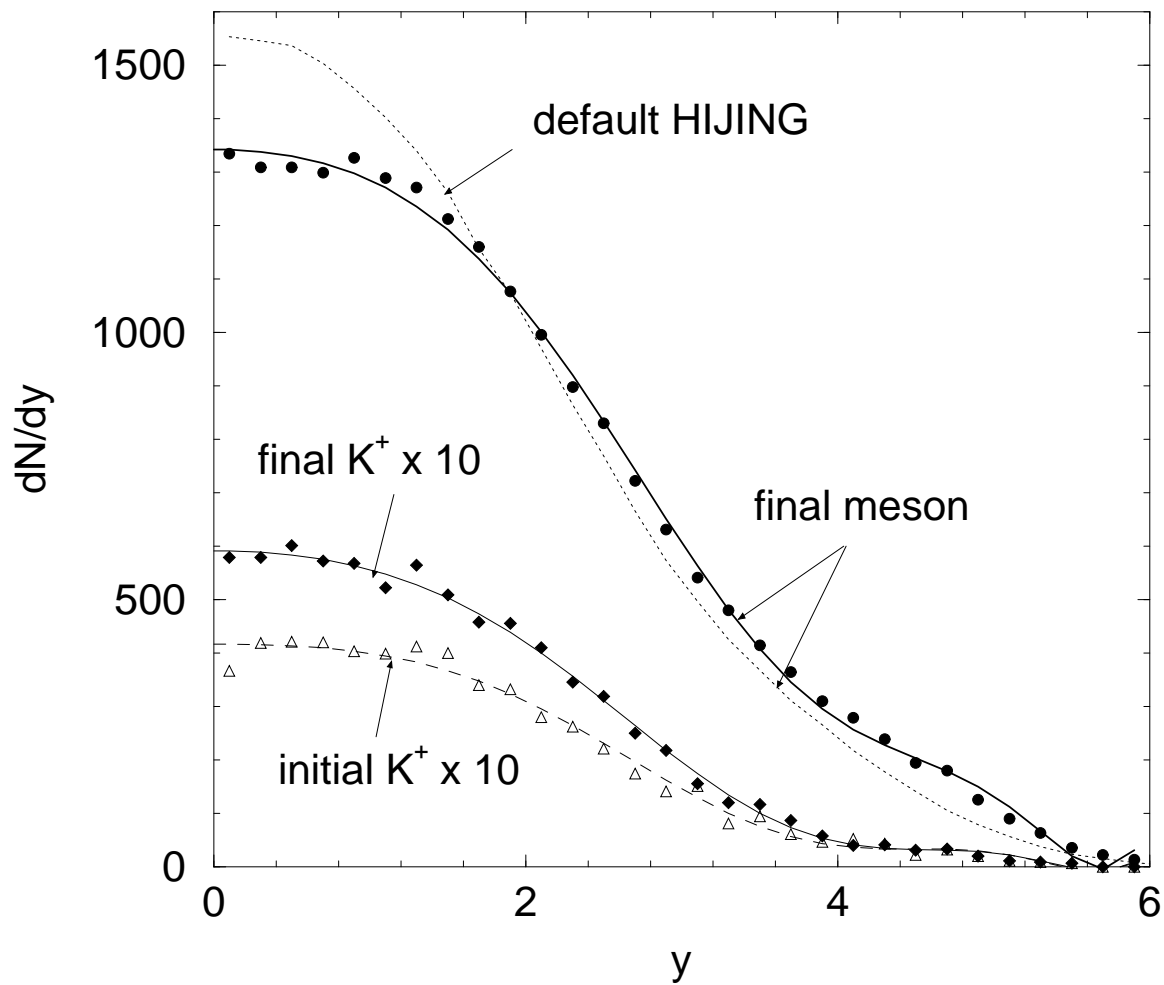,angle=270,width=6.0in}}
\caption{Meson rapidity distributions for central collisions.  
From Zhang, Ko, Li and Lin.} 
\end{figure}

\begin{figure}
\centerline{\epsfig{figure=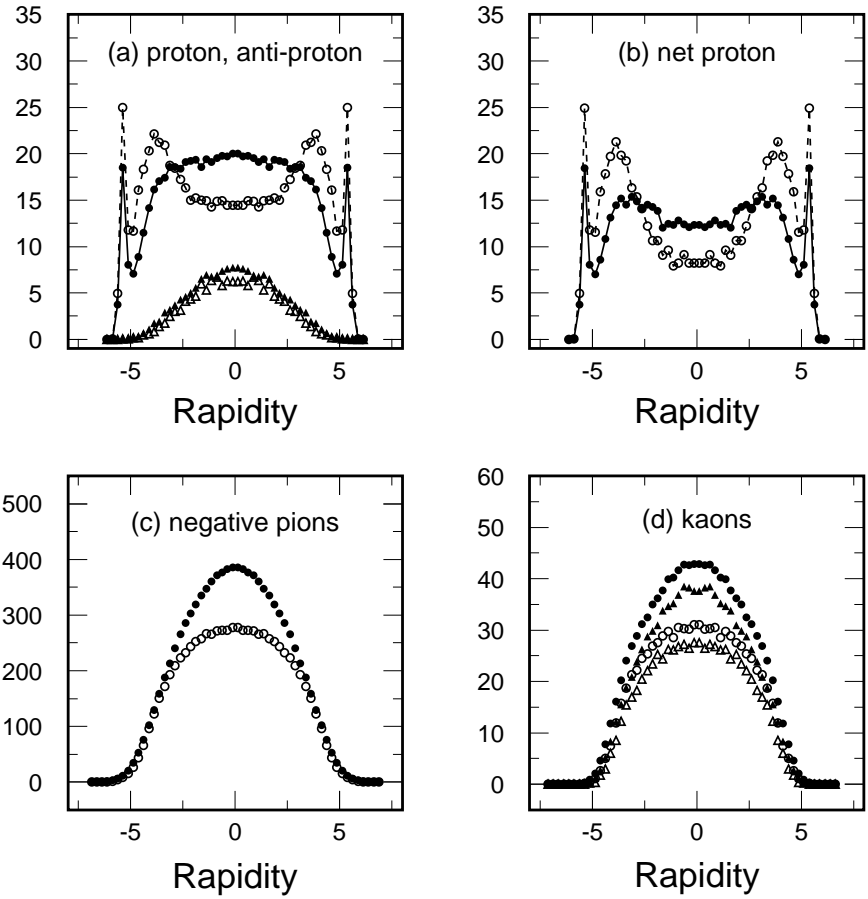,width=6.0in}}
\caption{Rapidity distributions averaged over collisions with $b < 3$ fm.
Full symbols include full rescattering, open symbols do not.  From Bleicher.} 
\end{figure}

\begin{figure}
\centerline{\epsfig{figure=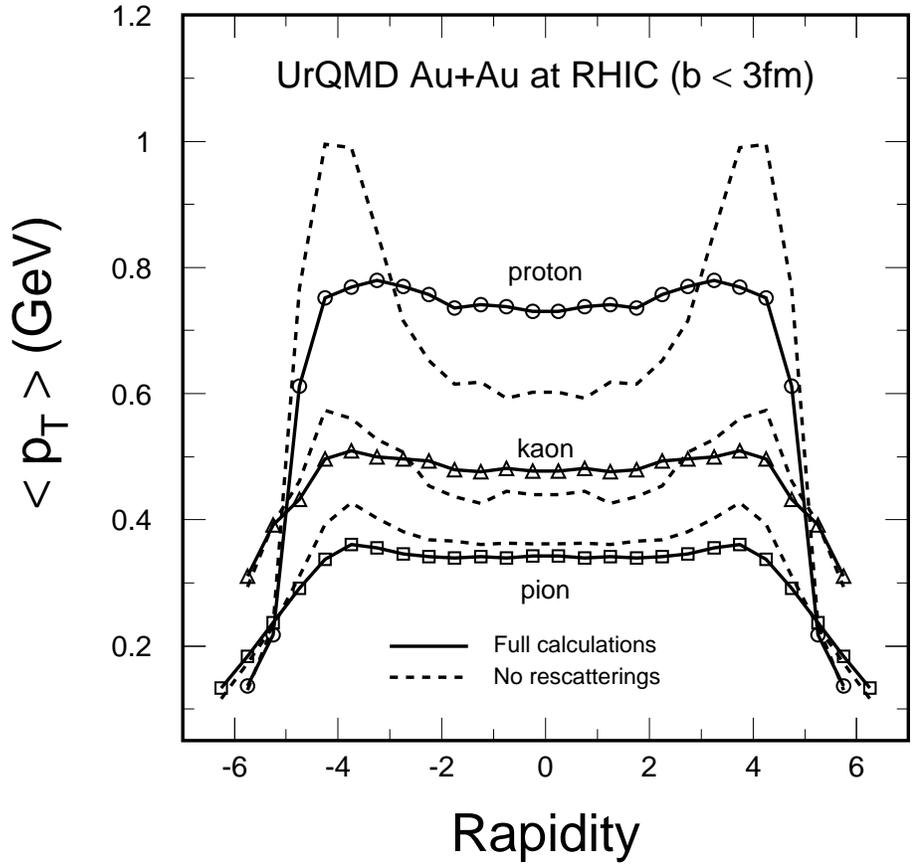,width=6.0in}}
\caption{Rapidity dependence of the average transverse momentum averaged
over collisions with $b < 3$ fm. Full symbols include full rescattering, 
open symbols do not.  From Bleicher.} 
\end{figure}

\begin{figure}
\centerline{\epsfig{figure=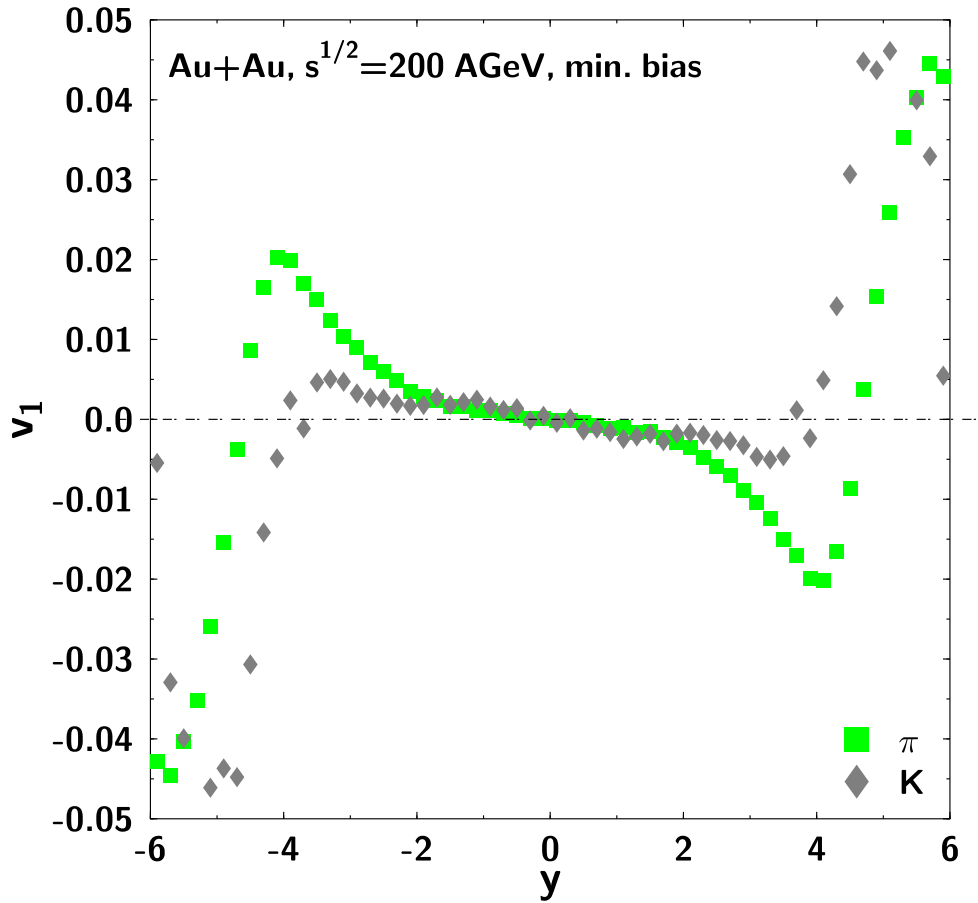,width=6.0in}}
\caption{Directed flow parameter as a function of rapidity.  From Bleicher.} 
\end{figure}

\begin{figure}
\centerline{\epsfig{figure=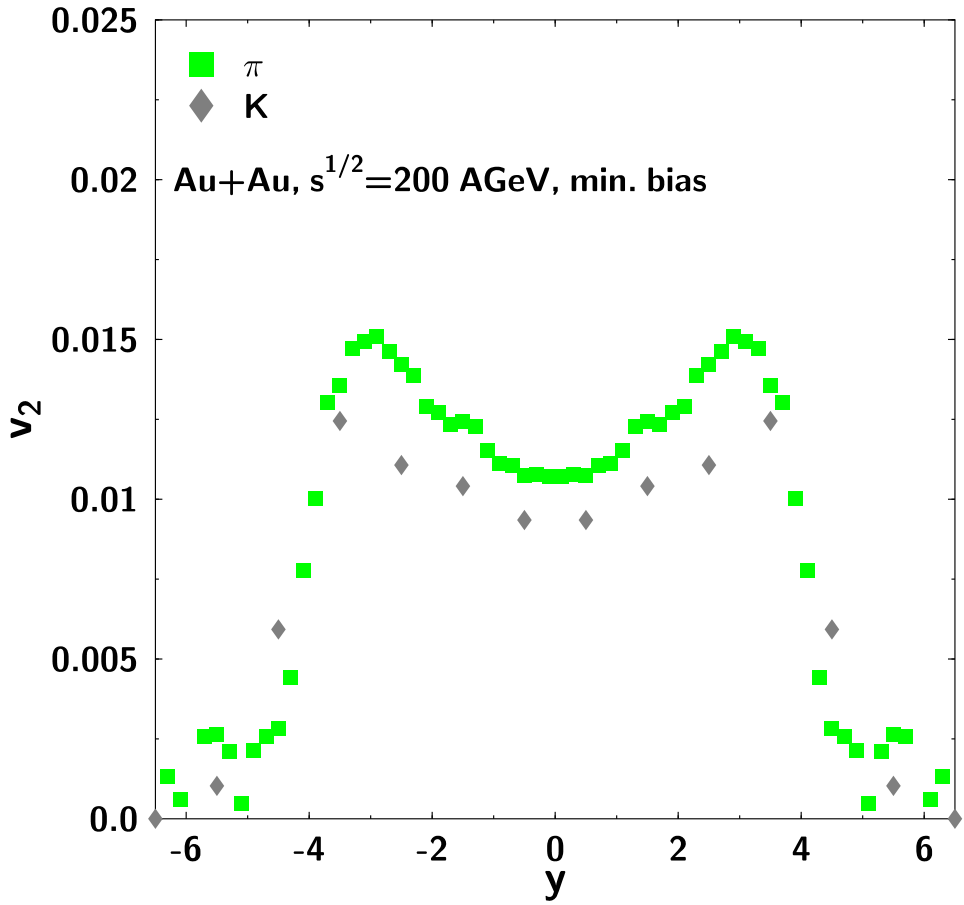,width=6.0in}}
\caption{Elliptic flow paramater as a function of rapidity.  From Bleicher.} 
\end{figure}

\end{document}